\documentclass[preprint2]{aastex}

\slugcomment{}

\shorttitle{Characteristics of NGC\,3516}
\shortauthors{Turner et al.}
\def\suzaku{{\em Suzaku}\ }

\begin{document}


\title{X-ray Characteristics of NGC\,3516: A View through the Complex Absorber} 

\author{T.J.Turner}
\affil{Department of Physics, University of Maryland Baltimore County, 
   Baltimore, MD 21250, U.S.A}

\author{L.Miller} 
\affil{Dept. of Physics, University of Oxford, 
Denys Wilkinson Building, Keble Road, Oxford OX1 3RH, U.K.}

\author{S.B.Kraemer}
\affil{Institute for Astrophysics and Computational Sciences, Department of Physics, 
The Catholic University of America, 
Washington, DC 20064}

\and 

\author{J.N.Reeves}
\affil{Astrophysics Group, School of Physical and Geographical Sciences, Keele 
University, Keele, Staffordshire ST5 5BG, U.K}

\begin{abstract}

We consider new {\it Suzaku} data for NGC\,3516 taken during 2009
along with other recent X-ray observations of the source.  The
cumulative characteristics of NGC\,3516 cannot be explained without
invoking changes in the line-of-sight absorption. Contrary to many other
well-studied Seyfert galaxies, NGC\,3516 does not show a positive lag
of hard X-ray photons relative to soft photons over the timescales
sampled. In the context of reverberation models for the X-ray lags,
the lack of such a signal in NGC\,3516 is consistent with flux
variations being dominated by absorption changes. The lack of any
reverberation signal in such a highly variable source disfavors
intrinsic continuum variability in this case. Instead, the colorless
flux variations observed at high flux states for NGC\,3516 are
suggested to be a consequence of Compton-thick clumps of gas crossing
the line-of-sight.

\end{abstract}


\section{Introduction}

Signatures of ionized gas are commonly observed in Seyfert galaxies. Well-studied  
sources reveal multiple zones of X-ray absorbing gas covering a range of ionization state, 
column density, covering fraction and kinematics 
\citep[e.g.][]{netzer03a,behar03a, kaspi02a,steenbrugge05a,blustin05a,mckernan07a,blustin07a}.  
The detection of deep K-shell absorption lines  from very highly ionized species of 
Fe \citep[e.g.][]{reeves04a,kaspi02a,turner08a} 
has shown that X-ray absorbing gas can be spectroscopically traced up to 
equivalent hydrogen column densities 
$N_{\rm H} \sim 10^{24} {\rm cm^{-2}}$. The possible existence of circumnuclear 
gas at high column densities is of great interest with regard to 
understanding the mass flow around the black hole 
and energetic considerations for the system. 
UV spectroscopy  unequivocally shows AGN to possess 
complex absorption systems and so  the presence of a complex of 
X-ray absorbers should be unsurprising. 

The systematic spectral hardening exhibited by Seyfert-type AGN as they  
drop in X-ray flux 
\citep[e.g.][]{papadakis02,pounds04a, pounds04b,vaughan04a, miller07a,miller08a,turner08a}  
can be modeled by changes in covering fraction of X-ray absorbing gas. There are a few cases 
where changes in individual lines 
 trace rapid changes in the X-ray absorber. For example, 
an  absorption line from Fe {\sc xxv} detected in NGC 3783 varies on timescales of 
 days \citep{reeves04a}, likely originating $\sim 0.1\,$pc from the nucleus. 
In  NGC~1365 variable absorption lines have been detected from  Fe\,{\sc xxv} and Fe\,
{\sc xxvi} supporting   
a picture in which the nucleus suffers variable obscuration by an absorber whose covering fraction changes 
on short timescales   \citep{risaliti05a,risaliti07a}. 

In the broad-line Seyfert 1 galaxy (BLSy1) NGC\,3516 (z=0.008836;
\citealt{keel96a}), X-ray data covering 0.5-10\,keV have revealed a 
strong signature from a  variable X-ray absorbing outflow. X-ray grating data from {\it Chandra} HETG and {\it XMM} RGS 
show discrete absorption features tracing a range of column densities, ionization states and velocity components for the 
 gas \citep[e.g.][]{turner08a,mehdipour10a}. 
Here we present new {\it Suzaku} data from NGC\,3516, 
taken during 2009 October. 
We also reconsider recent X-ray observations of this source, seeking to 
reconcile the spectral and timing behaviour and thus elucidate the true nature of the X-ray 
signatures of this AGN. 


\section{The Suzaku Observations}

Four co-aligned \suzaku \citep{mitsuda07} telescopes  focus X-rays on to 
CCD cameras comprising the X-ray Imaging Spectrometer \citep[XIS][]{koyama07}. 
XIS units 0,2,3 are front-illuminated (FI) while  XIS\,1 is a back-illuminated CCD. XIS\,1 
has an enhanced soft-band response but  lower area at 6\,keV than the FI CCDs as well as a larger 
background level at high energies.  XIS\,2 failed on 2006 November 9 and hence was not used.
\suzaku also carries a non-imaging, collimated  Hard X-ray Detector  
\citep[HXD,][]{takahashi07}, whose PIN detector provides useful AGN data typically over 
the range 15-70\,keV. 

A \suzaku  observation of NGC\,3516 was made 2009 Oct (OBSID 704062010) 
and the data  were  reduced using v6.9 of {\sc HEAsoft}. We screened the XIS and PIN events  
to exclude data during passage through the South Atlantic Anomaly, starting and
ending within 500 seconds of entry or exit.  Additionally we 
excluded data with an Earth  elevation angle less than 10$^\circ$ and 
 cut-off rigidity $>6$ GeV. The source was observed at the nominal 
center position for the HXD. FI CCDs were in $3 \times 3$ and $5 \times 5$ 
edit-modes, with normal clocking mode. We selected good  
events with grades 0,2,3,4 \& 6 and removed hot and flickering pixels using the 
SISCLEAN script.  The spaced-row charge  injection (SCI) 
was utilized. The exposure times were 222 ks per XIS unit and 178 ks for the PIN. 
XIS products were extracted from 
circular regions of 3.5\arcmin \, radius while background spectra were extracted from a region 
of radius 2.5\arcmin \, (offset from the source and avoiding the chip corners,
where there are data from the calibration sources). 

For the 2009 HXD PIN analysis we used the model ``D'' background
(released 2008 June 17 
\footnote[1]{http://www.astro.isas.jaxa.jp/suzaku/doc/suzakumemo/ suzakumemo-2007-01.pdf}).  
The time filter resulting from screening the observational data was applied to the background events 
model.   
The ftool {\sc hxdpinxbpi} was used to create a PIN background spectrum from the screened background data.   
 {\sc hxdpinxbpi} takes account of the 
form and flux level of the cosmic X-ray background 
\citep{boldt87,gruber99} in the  
{\it Suzaku} PIN field of view.  We used the response file  \\ 
\verb+ae_hxd_pinxinome3_20090826.rsp+ for spectral fitting. 

During the 2009  \suzaku observation NGC\,3516  was found to have source count rates 
0.507 $\pm 0.001$ (summed XIS 0,3 over 0.75-10\,keV)  and $0.049 \pm 0.002$ (PIN over 15-50\,keV) ct/s.
The background was $5\%$ of the total XIS count rate in the full band. 
The source comprised 9\% of the total counts in the PIN band. 
The source fluxes for recent observations of NGC\,3516 are shown in Table~1. 

Spectral fits utilized data  from {\sc XIS} 0 and 3,
in the energy range $0.75-10$\,keV
and from the {\sc PIN}  over $20-50$\,keV. 
Data in the range 1.75-1.9\,keV were also excluded from the XIS spectral analysis owing to 
uncertainties in calibration around the instrumental Si K edge. 
We note that data can be used down to lower energies, $\sim 0.3$\,keV, 
for timing analysis.  XIS\,1 was not used owing to the 
higher background level at high energies. 
XIS  data were binned at the HWHM resolution for each instrument, optimal for detection 
of spectral features, while PIN data were binned to be a minimum of 5$\sigma$ 
above the background level for the spectral fitting.  
In the spectral analysis, the PIN flux was increased by a 
factor 1.18 for the new data, appropriate for the cross calibration 
of instruments for the 2009 observation 
 \footnote[2]{ftp://legacy.gsfc.nasa.gov/suzaku/doc/xrt/suzakumemo-2008-06.pdf}.
The new data were considered in conjunction with 
those  from 2005, whose screening and data reduction are described by 
\citet{markowitz08a}. 

\section{Spectral Fitting Results}

{\it Chandra} images of NGC 3516 \citep{george02a} have shown an extended 
soft X-ray emission component contributing an observed flux  
$F_{\rm 0.4-2\,keV} \sim 10^{-14} {\rm erg\, cm^{-2}\, s^{-1}}$ within 
$\sim 2$  kpc (assuming 
$H_o =73\, {\rm km\, s^{-1}\, kpc^{-1}}$ throughout). In addition, there is 
a known off-nuclear X-ray source, 
CXOU 110648.1+723412 \citep{george02a} that cannot be separated from
the nuclear emission using {\it Suzaku}. 
However these two provide only a small fraction of the 
total soft-band flux for the epochs considered here. 

\begin{figure}
\epsscale{.75}
\plotone{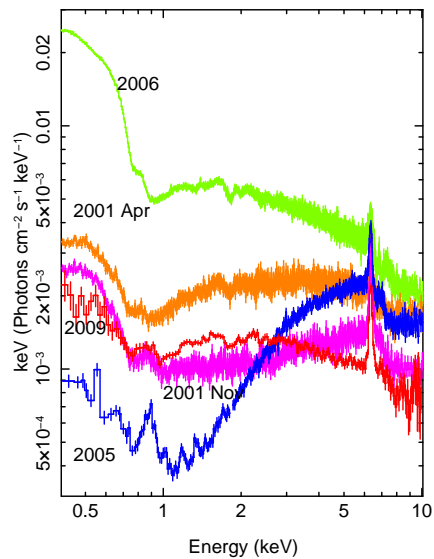}
\caption{The recent X-ray states of NGC\,3516 from {\it XMM} observations in 
2001 and 2006 and from {\it Suzaku} observations during 2005 and 2009.}
\label{fig:all}
\end{figure}

\begin{figure}
\epsscale{.75}
\plotone{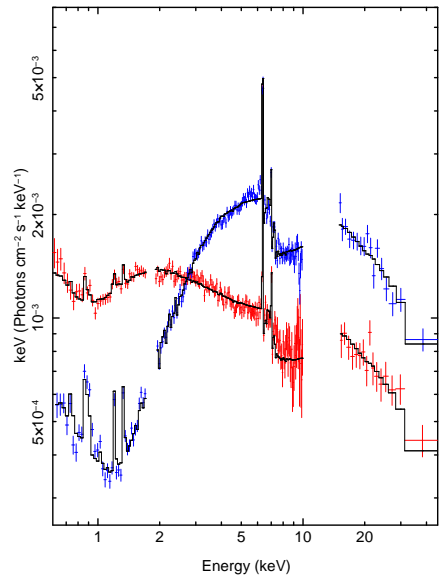}
\caption{\suzaku XIS and PIN data from 2005 (blue) and 2009  (red) along with 
the absorption model (black solid line).}
\label{fig:2suz}
\end{figure}

For an immediate visual comparison of data from recent X-ray
observations of NGC 3516, we plotted the 2009 XIS spectral data
with the 2005 {\it Suzaku} observation and other 
observations made using {\it XMM}  overlaid (Figure~\ref{fig:all}).  This
comparison reveals that during 2009 {\it Suzaku}
found NGC\,3516 in a relatively low flux state. 
 The data show that the line-of-sight absorption is obviously changing 
between epochs for NGC 3516. 
Although the hard X-ray flux levels are very similar in the 2005 {\it Suzaku} and
2001 Apr {\it XMM} data, the former shows a large amount of extra
absorption compared to the latter. The lowest level of hard X-ray flux is observed in the new data from 2009, although at this epoch 
the absorption appears to be also relatively low. 
It is clear that the source cannot be simply described by a  response in ionization of the absorbing gas 
to changes in the continuum flux.

\subsection{Fitting Absorption Models} 
\label{absfit}

From detection of discrete absorption lines and from 
broad spectral curvature, \citet{turner08a} found evidence for 
a multi-zoned X-ray absorber  in a joint  {\it XMM}/{\it Chandra} campaign during 2006. 
\citet{markowitz08a} also found complex absorption to be required to fit the spectrum of 
NGC\,3516 observed by {\it Suzaku} during 2005 (and in that spectral model, 
additionally required a modestly broadened component of Fe\,K$\alpha$ emission).  

We fit the {\it Suzaku} 2005 and 2009 data over 0.75-50.0\,keV using
{\sc xstar} v2.2 to parameterize the ionized gas. The 
warm absorber table was created with 
effective micro-turbulence velocity dispersion $\sigma=100$\,km\,s$^{-1}$ 
for gas of density $n=10^{10}\, {\rm cm^{-3}}$. 
Here we use the $\xi$ form of ionization parameter, 
$\xi = L/nr^2$, where $L$ is the 1-1000 Rydberg luminosity 
and $r$ the absorber-source distance.  

The spectral model was constructed with a powerlaw continuum, adding
zones of ionized gas until the data were parameterized adequately. The
photon index, column densities and ionization states of the gas were
allowed to be free but were initially linked to be the same values
during 2009 and 2005. 
 
 We assumed the gas zones to be layered over each other, with the highest column density zones 
closest to the nucleus: for this reason,  the covering fractions do not total 100\% (Table~2). 
The model included a full screen of neutral gas fixed at the Galactic line-of-sight 
column, $3.45 \times 10^{20} {\rm cm^{-2}}$, covering all components 
and parameterized using the {\sc xspec} model {\sc tbabs}  \citep{kalberla05a}. 

The fit yielded 
an underlying  photon index  $\Gamma=2.22\pm0.04$ covered by  zones of 
intrinsic absorbing gas  (Table~2) 
having three different covering factors.  
The outermost layer is Zone 1, comprising low-ionization gas, as  
 established in NGC~3516 from {\it Chandra} grating spectroscopy \citep{turner06a} and 
from  UV spectroscopy \citep{kraemer02a}. HETG 
spectra required two gas components to account for the soft-band 
absorption features measured during 2006 \citep{turner08a}. For the joint {\it Suzaku} fit 
we also found that the fit was significantly improved 
($\Delta \chi^2=63$) by allowing the low-ionization zone to have two components 
(Table~2 Zones 1a and 1b, having the same covering fraction). 
 The total column density for Zone 1 is consistent with the 
sum of the UV absorbing components  measured at other epochs \citep{kraemer02a}.  

Additional absorbing zones were found (Table~2) with column densities 
 $\sim 2 \times 10^{23} {\rm cm^{-2}}$ and $\sim 2 \times 10^{24} {\rm
  cm^{-2}}$, both having log $\xi \sim 2$. The covering fraction of
the Compton-thick zone is around $70\%$ in 2005 and 2009 spectra. We note that this 
hard component 
can, alternatively,  be modeled as reflection, as shown by \citet{markowitz08a}.

{\sc xstar} does not calculate the scattered component of X-ray flux
from the absorbing gas, however it does calculate the re-emitted line
spectrum and we included such a component in the model, placing it 
behind the Zone 1 absorber complex. 
With such a  placement the re-emission 
spectrum 
accounts for the strength of Fe K$\alpha$ emission without over-predicting 
the soft line emission (that is  somewhat suppressed by absorption). 
The ionization parameter 
of the emitting gas was left free to encompass the possibility that 
emission arises in a gas component other than one of the absorbers and indeed, 
the fitted ionization parameter was 
 found to be $\xi=1.53\pm0.10$, different to  
the absorbing gas.  The fit gave a reduced 
$\chi^2_r=1.24/358$ degrees of freedom (d.o.f.). 
Interestingly, the emission spectrum 
fitted here is consistent with the ionization-state found for soft-band
lines in {\it XMM} RGS grating data \citep{turner03b}.  

We stress that
the absorption model does not over-predict the Fe K$\alpha$ flux, or
the flux of any emission line, even if the gas zones are assumed to
fully cover the continuum source.
The strength of the emission from the absorbing gas  depends on 
the gas geometry, column density and  covering/filling factors. 
Even large column densities for the absorber 
can result in a small predicted line equivalent width 
\cite{miller09a,yaqoob10a} when there is expected to be a large opacity at the line energy. 
It is not suprising that the absorber emission has not been isolated in these data, given 
 the gas geometry suggested in this case.   

In this absorption model, the covering fractions of the gas layers varied  
significantly between epochs (Table~2). 
Importantly, Zone 2  had a  larger covering fraction during 2005 (99\%) than during
2009 (86\%), despite the hard-band continuum flux being a factor of 2 higher (Figure~\ref{fig:all}).  

The most highly ionized zone detected at any epoch is that traced by 
 absorption lines from Fe {\sc xxv} and Fe {\sc xxvi}  whose depths indicate an origin in 
gas having $N_{\rm H} \sim 3 \times 10^{23} {\rm cm^{-2}}$ and log\,$\xi \sim 4.3$ \citep{turner08a} 
These lines were detected in  {\it XMM} and {\it Chandra} data from 2006, and in  {\it Suzaku} data
from 2005,  however,  they were
not detected during the {\it Suzaku} observation of 2009, with limits on the
equivalents widths of (narrow) Fe {\sc xxv} and {\sc xxvi} absorption 
found to be $ < 5$\,eV and  $< 20$\,eV respectively.

The absorption modeling and covering fraction changes are fundamentally consistent with the 
conclusions of  \citet{markowitz08a} and \citet{turner08a}: i.e. the source has a complex absorber and 
changes in covering by a high column of gas $\sim 10^{23}{\rm cm^{-2}}$  of 'intermediate' ionization (in terms of the X-ray absorbing zones detected in NGC 3516)  dominate the spectral variability. 
Small differences between the fit of \citet{markowitz08a} and that presented here arise because 
we are  
fitting under the 
joint constraints from 2009 and 2005 epochs, and  also because we utilize a new version of {\sc xstar}  
and use a slightly different model construction. 

Constraining the covering fractions of the gas to be the same for 2005 and 2009, allowing 
only the column densities and ionization states of the layers to be free for each epoch, 
does not provide a satisfactory joint fit to the two epochs ($\chi^2_r=4.8/355\,$ d.o.f.). 
It is clear that the source cannot be parameterized satisfactorily 
without invoking such a mode of variation. 

Small improvements to the fit can, of course, be obtained by allowing
the absorber column densities and ionization-states to also vary
between epochs. However, as {\it Suzaku} cannot distinguish which of
the layers of gas are varying, we did not find it useful to pursue
this line of consideration further.

\subsection{Fe\,K$\alpha$ Line Variability}

To test for variability in the Fe\,K$\alpha$ line emission we re-fit
the 4.0-7.5\,keV data using a simple absorbed power-law model
for the local continuum.  The line was modeled using a  simple Gaussian
profile. Initial fits found the line width to be 
consistent with the HEG value $\sigma=40$\, eV, at all 
epochs \citep{turner08a} and so 
the fits were rerun with the line width fixed at that  value. 

The joint 2005/2009 {\it Suzaku} fit 
yielded a peak energy for the line E$=6.40 \pm 0.06$\,\,keV. The
normalization of the Fe\,K$\alpha$ line was $5.75\pm 0.24 \times
10^{-5}$\,photons\,cm$^{-2}$\,s$^{-1}$ (an equivalent width of 164\,eV
against the observed continuum) during 2005 and $4.16^{+0.33}_{-0.32}
\times 10^{-5}$\,photons\,cm$^{-2}$\,s$^{-1}$ (equivalent width
253\,eV) during 2009. Thus comparison between the two mean spectra observed using {\it Suzaku} 
reveals  strong evidence ($>99.9$ \% confidence) for 
variability in Fe\,K$\alpha$ line flux over timescales of years.  

To investigate the line behavior on shorter timescales, the 2001 Apr, 2001 Nov and  
2006 {\it XMM} data were re-fit with the new model. Fe\,K$\alpha$ line fluxes
are detailed in Table~3. In addition to these fits, the individual OBSIDs comprising the 
{\it XMM} data for 2006 were fit individually, yielding Fe\,K$\alpha$ fluxes 
sampled on timescales of days: however, no significant variability was detectable 
on such a timescale (Table~3). 

\begin{figure}
\epsscale{1.0}
\plottwo{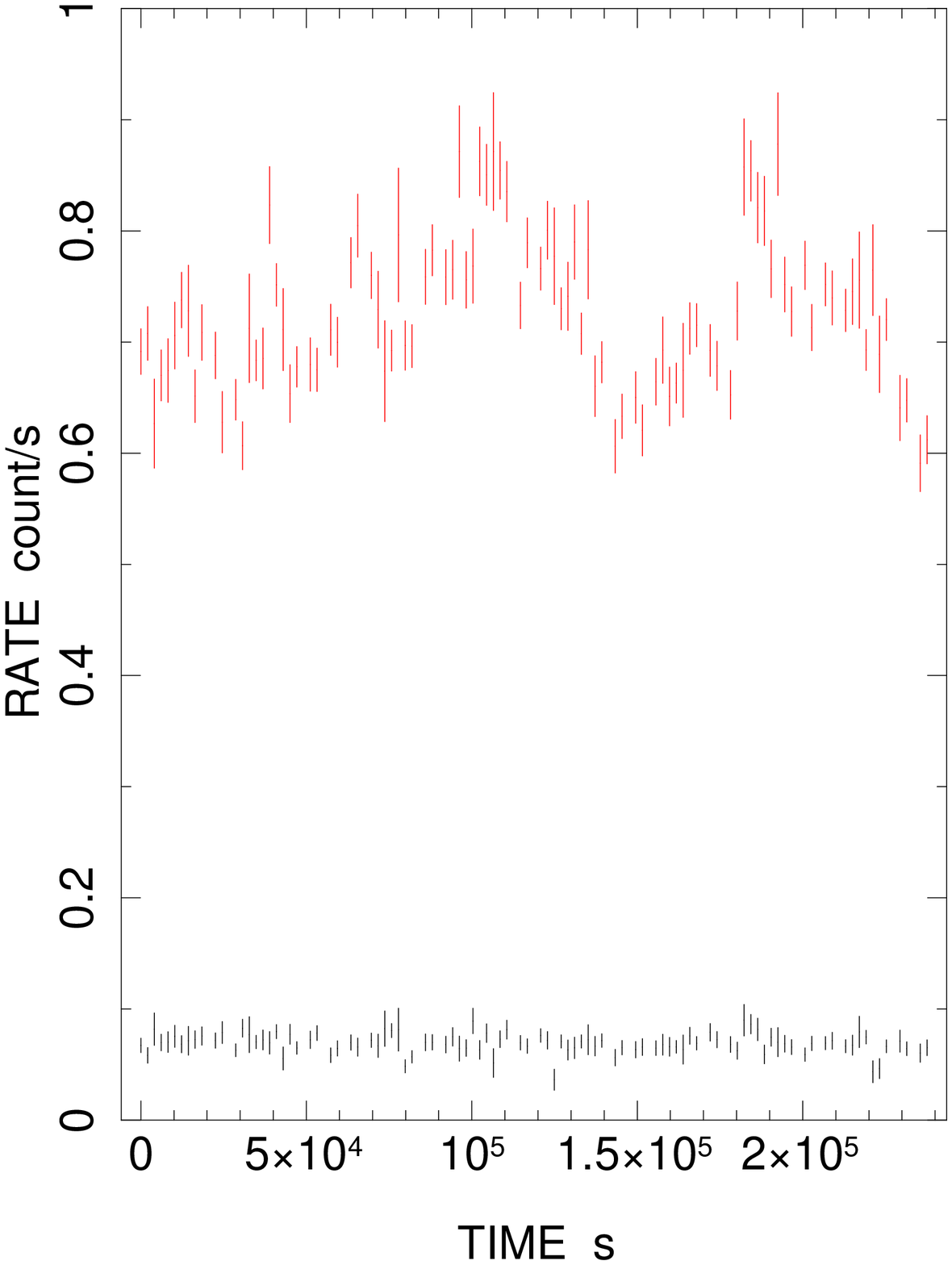}{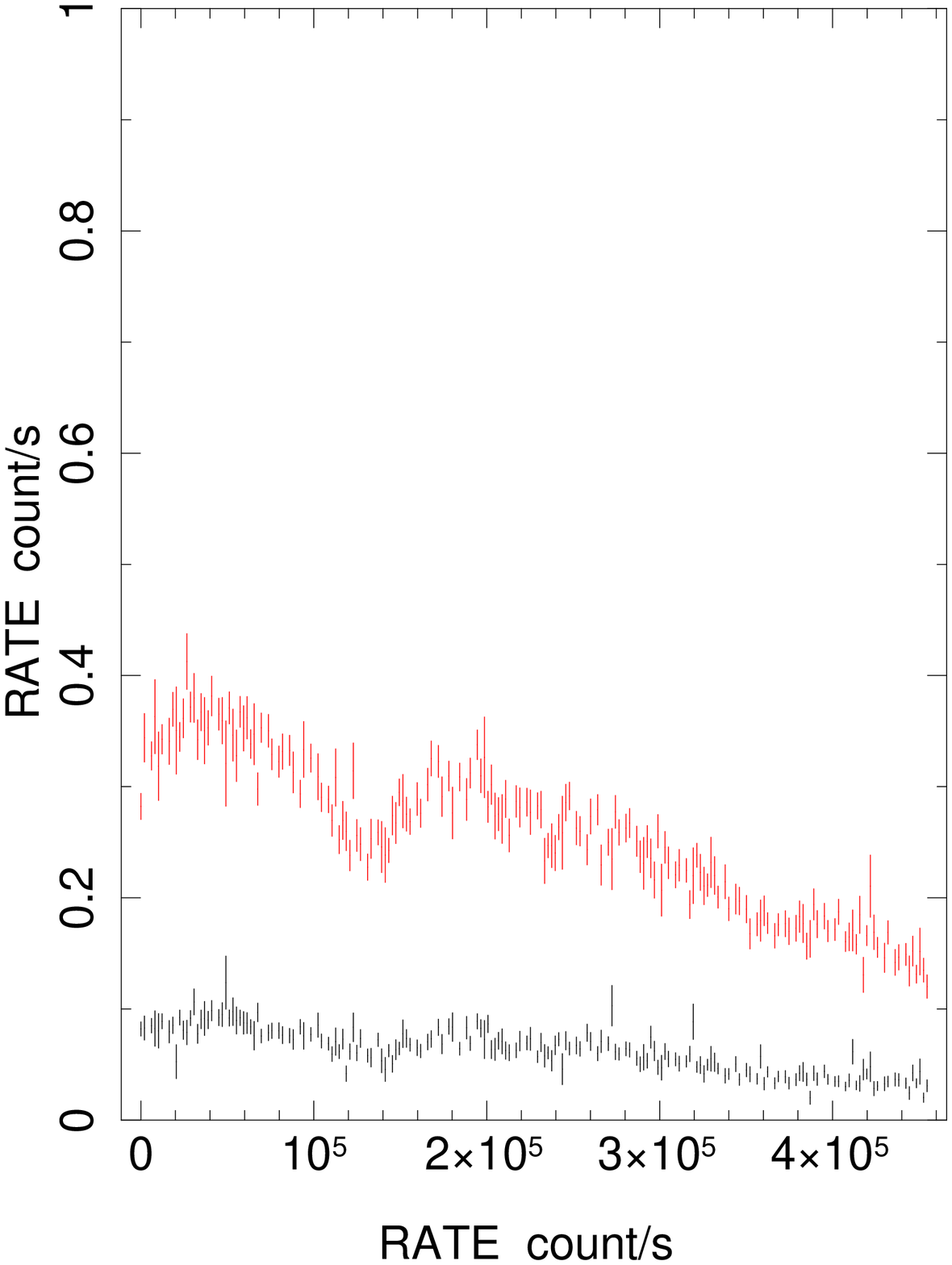}
\caption{\suzaku XIS$0+3$ time series in 2048s bins, from the 2005 and  2009 Suzaku 
observations. Red points represent 4-7.5\,keV and black the 0.3-1\,keV band.}
\label{fig:suzlc}
\end{figure}

\section{Timing Analysis}

We examined the hard (4.0 - 7.5\,keV) and soft-band (0.3-1\,keV) light
curves from the {\it Suzaku} observations. As was evident from the
spectral analysis (Figure~\ref{fig:2suz}), the hardness ratio is very different for the two
epochs (Figure~\ref{fig:suzlc}).

\begin{figure}
\begin{center}
\resizebox{65mm}{!}{
\rotatebox{270}{
\includegraphics{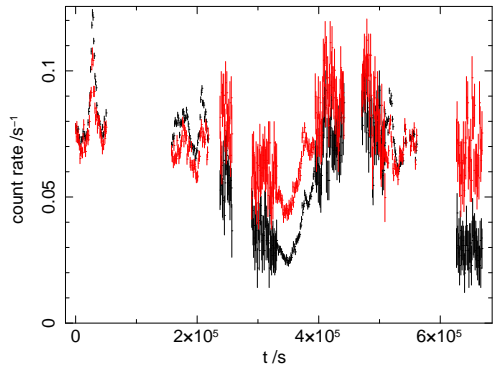}}}
\end{center}
\caption{The light curve in 2048s bins from 2006 XMM pn and {\it
    Chandra} HETG data with soft (black: 0.3-1.0\,keV) and hard bands
  (red: 4.0-7.5\,keV) overlaid.  The {\it XMM} light curve rates have been renormalized
  to the {\it Chandra} count rate, and HETG points can be recognized
  most easily by the relatively large uncertainties on the count rate. }
 \label{fig:2006lc}
\end{figure}

The best constraints on the 
timing properties of NGC\,3516 are currently offered by the combined data of 2006. At that epoch the
source was bright and, taken together, the  {\it XMM} and {\it Chandra} exposures
 provide a long baseline of coverage 
\citep{turner08a}. We re-constructed the combined  {\it XMM}/{\it Chandra} light curve 
 in selected energy bands 
(Figure~\ref{fig:2006lc}).  The effective areas and spectral responses of the
{\it XMM} and {\it Chandra} data differ: to create a single light curve, 
the expected {\it XMM} and {\it Chandra} count rates in narrow energy bins were 
calculated from the mean spectral model. Next, the time-dependent observed
{\it XMM} count rate in those energy 
bins was rescaled to the expected {\it Chandra} count rate using that mean conversion
between instruments.  The light curve in broader energy bands was then calculated by summing
the renormalized values, propagating {\it XMM} statistical errors, taking account of the
energy-dependent re-normalization.  By scaling to the {\it Chandra} instrument the 
uncertainties on the {\it Chandra} part of the light curve remain at their Poisson values, whereas
on the {\it XMM} part the uncertainties are slightly worse than Poisson, albeit still much
smaller than the {\it Chandra} uncertainties.  By re-normalizing in fine energy bins, to
first-order the effect of spectral variations on the accuracy of the cross-instrument
combination are minimized.

For a quantitative assessment of the source variability behavior we 
follow \citet{miller10a, miller10b}, in which a maximum-likelihood
method is used to measure both the power spectral densities (PSD) and the
cross-spectral densities of time series created in a number of broad bands of photon energy.
We assume that the physical process(es) generating the count rate variations in the time series  
is Gaussian (equivalent to  the 
Fourier transforms of the intrinsic time series, before being observationally
sampled, having uncorrelated phases).  The PSD and cross-spectral density are defined
as the Fourier transforms of the autocorrelation and cross-correlation functions.   
To find the best-fitting PSD and cross-spectra we compute the expected covariance matrix
between time samples for an initial set of PSD and cross-spectral
values in bins of frequency and iterate to a maximum likelihood solution
using the method of \citet*{bond98a}.
The covariance matrix includes the effects of shot noise.   
The time lags between bands are evaluated as parameters in the cross-spectral density
\citep{miller10a, miller10b}.

\begin{figure}
\begin{center}
\resizebox{60mm}{!}{
\rotatebox{270}{
\includegraphics{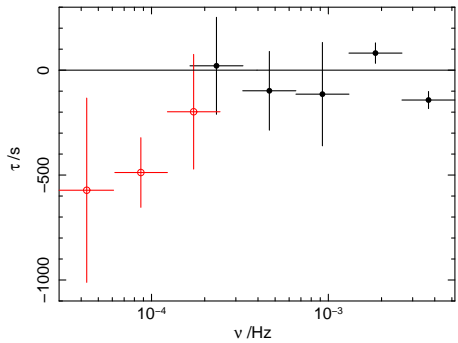}}}
\end{center}
\caption{Lag spectra for 2006 XMM data, calculated for 4-7.5\,keV versus 0.3-1\,keV. 
On this plot, positive values of $\tau$ 
would represent a lag of hard photons relative to the soft.} 
 \label{fig:lags}
\end{figure}

\begin{figure}
\begin{center}
\resizebox{60mm}{!}{
\rotatebox{0}{
\includegraphics{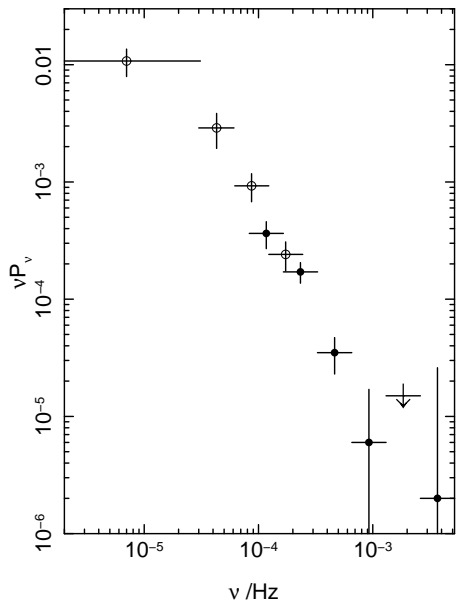}}}
\end{center}
\caption{PSD from 2006 data over 0.3-1.0\,keV. The open points denote the 
combined {\it XMM} and {\it Chandra} data using 
2048 s bins, the solid points denote the {\it XMM} data alone, 
sampled using 96 s bins.} 
\label{fig:psd}
\end{figure}

We applied these methods to the 2006 combined {\it XMM}/{\it Chandra} time series; above 
$10^{-4}$\, Hz the {\it XMM} data were used alone, owing to the low signal-to-noise ratio 
for {\it Chandra} data sampled at those frequencies.  
This new evaluation of the PSD (Figure~\ref{fig:psd}) confirms previous conclusions 
\citep{edelson99a} that the PSD shows a steep drop to high frequencies.
In contrast to the long {\it XMM}/{\it Chandra} campaign of 2006, the {\it Suzaku} data 
yielded a relatively short baseline, relatively low amplitude 
of variability and did not yield conclusive results for the lag spectra. 
Intriguingly the lag spectra for 2006 data from NGC\,3516 showed no positive time lags
(Figure~\ref{fig:lags}). At the lowest frequencies sampled there is weak evidence
for negative time lags (soft band lagging the hard band) but as the timescales concerned
are approaching the longest periods sampled by these observations, caution is required
in their interpretation, and independent observations are required to confirm their
reality.

\section{Discussion} 
\label{sec:discuss}

\subsection{Spectroscopy Results} 

Comparison of recent X-ray spectra of NGC~3516 shows that 
the source has a relatively hard spectrum at some epochs
of high X-ray flux. A review of historical X-ray observations of NGC\,3516
confirms that there have been numerous epochs during which the source has shown
heavy absorption by low-ionization gas, often at 
high levels of continuum flux \citep{kruper90a,ghosh91a,kolman93a,guainazzi01a}.  
The covering changes for the absorbing gas are not well-correlated with the 
observed X-ray continuum flux in NGC\,3516, implying that, in addition to the flux variations that are a direct 
result of obscuration changes, one must invoke either intrinsic continuum variations or 'colorless' 
variability originating from the passage of Compton-thick blobs of gas across the line-of-sight. 
Reverberation results 
favor the latter possibility  (see Section~\ref{sec:timing_discuss}) and in that case 
the  additional component of the model forms a natural extension to the absorber spectral model.  

The light-bending model of \citet{miniutti04a}, which attempts to account for flux and
spectral variability in AGN by varying the distance of the continuum
source perpendicularly from the accretion disk,
cannot naturally account for behavior where the
source spectrum is found to be hard at a high flux state. Adherence to
this model would require an additional variability mechanism to account for
all of the X-ray states observed in NGC\,3516. Invocation of another
source of continuum variability would require an explanation of the
lack of positive lags between hard and soft bands, which are observed
in other AGN.

Other models that have been suggested to explain the spectral
variability of NGC\,3516 are also not favored by this
comparison of spectral states. \citet{netzer02a} found {\it ASCA}
observations of NGC~3516 to be consistent with changes in the
ionization state of the X-ray absorber in response to variations in
the continuum flux level.  \citet{mehdipour10a} also favor the
`responsive gas' scenario (although based upon a column density an
order of magnitude lower than that fitted by \citealt{turner08a}, owing
to their different modeling of the X-ray continuum and, in particular, the
inclusion of a black body component that we do not require). However, a responsive-gas model
cannot explain the behavior of NGC\,3516 across all the X-ray
observations; variable-covering absorption is required to explain the 
overall behavior of the source. 
In other work, \citet{bhayani10a} have claimed that 
differences in the {\sc RMS} variability 
between the 4.6-5.0\,keV band and 5.0-5.4 and 5.4-5.8\,keV bands for the 2001 {\it XMM} data 
disfavor an absorption model. However, the predicted 
measure of {\sc RMS} variability in a given band 
depends on the column and ionization-state of the absorbing layer that dominates the variations
in that band.  As the absorption model matches the X-ray spectrum of NGC\,3516
at all flux levels and time slices, it follows {\em de facto} that the model reproduces also the 
{\sc RMS} spectral variability.  

Of course the partial-covering  model does not rule out a response of
the absorbing gas to changes in the incident flux: 
it is  inevitable (and therefore predicted) that covering changes for inner zones of a
complex absorber modify the flux and spectral shape of the continuum reaching zones further out, possibly
leading to measurable changes in the ionization balance of the gas.  
 
As discussed for the 2006 {\it XMM}/{\it Chandra} data 
\citep{turner08a}, numerous absorption 
lines are detected in the HETG and RGS gratings, providing a firm basis for the 
absorption model in NGC~3516. 
The combined signature of multi-zoned X-ray absorbers produces
curvature in the observed spectral shape, while spectral variability can be explained by 
changes in covering fraction of the layers.  
Although current  X-ray data do not 
allow the gas covering fraction to be measured directly from the absorption lines, 
this quantity can be constrained from the broad X-ray spectral curvature. 

This X-ray absorption model  comprises 
a compelling extension of the well-studied UV absorption systems. 
NGC\,3516 shows some of the strongest and most variable 
intrinsic blueshifted UV absorption lines of any Seyfert 1 galaxy \citep{kolman93a,goad99a}.
The C {\sc iv} absorption is particularly deep and broad 
and has led some to compare the systems in NGC\,3516 to those in 
broad absorption line QSOs \citep{ulrich88a}. 
UV absorption features arise in multiple zones having  covering 
fractions $< 1$ \citep{kraemer02a}.   \citet{kraemer02a} 
resolve several kinematic components of absorption in the UV lines. Those authors  also 
isolated zones of gas having signatures spanning both the UV and X-ray 
regimes, confirming an association between the two bands 
that had been previously suggested \citep{guainazzi01a}.  
Estimates placed the C {\sc iv} 
origin at $\sim 10^{16} {\rm cm}$  \citep{koratkar96a} 
with other lines arising out to $\sim 10^{18} {\rm cm}$ from the central source \citep{kraemer02a}. 


Similarly to Seyfert type\,2 AGN, NGC\,3516 shows a bipolar morphology for
the narrow-line-region gas \citep{pogge89a}.  In  NGC 4151, 
 {\it STIS} data show  evidence for fast-moving clouds
existing over a wide angle and originating close to the nucleus \citep{hutchings99a,crenshaw07a}, 
possibly collimating the nuclear 
radiation that excites the NLR gas \citep{kraemer08a}. A similar   model may apply to NGC\,3516, where   
a toroidal distribution of outflowing clouds might
collimate the nuclear radiation. 
Such a toroidal gas distribution could 
provide  X-ray absorption and emission features from 
different zones within the flow. If our view of NGC\,3516 intercepts the edge of the collimating
flow we might naturally 
expect the source behavior to be more variable than sources viewed at face-on or edge-on 
orientations.

In the context of the partial-covering model, variations of the Fe\,K$\alpha$ flux  
may arise if the line-emitting region suffers variable obscuration. However, the line flux also 
depends on the illuminating flux reaching the emitting gas  
and so the line behavior might be expected to appear complex in a multi-layer absorber model. 

HEG data find the Fe\,K$\alpha$ emitter to have a bulk velocity consistent
with zero \citep{turner08a}. The HEG constraint on the line width is 
$25 < \sigma < 50$\,eV, equivalent to FWHM velocity broadening in the range
$3000 - 6000$\, km/s.  For gas in a Keplerian orbit the line width
corresponds to an origin (4-17)\,sin$^{2}i$\,light-days from the
nucleus (where $i$ is the angle of inclination of the orbital plane, assuming 
a black hole mass $2.95 \times 10^7 {\rm M}_{\odot}$, \citealt{nikolajuk06a}). 

\subsection{Timing Results} 
\label{sec:timing_discuss}

There has been mounting support for the idea that large amplitude and
rapid X-ray flux variations  are caused by changes 
in the line-of-sight absorption \citep{boller97a,
 guainazzi98a,tanaka04a}.  In such a picture, the low/hard states of AGN 
may be attributable to  increases in X-ray obscuration, as  
suggested for NGC\,3516 based on {\it BeppoSAX} and 
{\it ASCA} data \citep{nogami04a}.  Deep dips in X-ray flux have been traced 
by the light curves of 
 MCG-6-30-15 \citep{mckernan98a} and NGC\,3516
\citep{turner08a}, where data allow us to track the 
 apparent ingress through to the egress of an occultation event. 
Interestingly, the detailed dip profiles  match the profile of  dip events in some 
  X-ray binaries, e.g. GRO 1655-40 \citep{tanaka03a}, where in that case the dip 
 was attributed to an increase 
in covering fraction of a complex X-ray absorber. 
Spectral variability in Cir X-1 has also been attributed to changes in covering of a complex 
absorber \citep{brandt96a}.  The 
deep dips in NGC\,3516 and MCG-6-30-15  are similarly consistent with occultation events. 
The large number of similarities in observed timing and spectral properties of 
AGN and stellar binary systems suggests that 
complex and variable absorption dominates the observed X-ray properties for accreting systems over 
 stellar to galaxy size scales.  This, in turn, suggests that the often-outflowing absorber is 
closely related to the 
accretion process, and therefore of the highest importance in understanding the fundamental 
mechanisms in these systems.  

Close study of key sources is required to decouple intrinsic continuum variations from 
absorption effects. The 2006 {\it XMM}/{\it Chandra} campaign
\citep{turner08a} showed that in the 0.5-10\,keV band, there appear to
be two characteristically different modes of X-ray variability: 
one which results in opacity changes and thus changes in hardness ratio, the
other, particularly at high fluxes, which results in no change in hardness ratio.
In the context of the complex absorption model, the latter mode may apply when
the continuum is uncovered by the fitted absorbers: in the 2006 data this
occurred above a flux 
$F_{2-10} \sim 3 \times 10^{-11} {\rm erg\, cm^{-2}\, s^{-1}}$.  
Variability above this threshold was attributed 
either to intrinsic variations in the continuum or to  changes in covering
by Compton-thick 'bricks' \citep{turner08a}.

Recent long X-ray exposures on two bright AGN have allowed significant
progress with regard to our general understanding. {\it Suzaku} data from NGC 4051 show that the
frequency- and energy-dependence of the observed time lags can be
explained by the effects of reverberation from a thick shell of material 
extending to $\sim 600$ gravitational radii from the continuum source 
\citep{miller10a}. That work was the first interpretation of the 
previously-known positive AGN time 
lags, that dominate the lag spectra, as arising from reverberation within the X-ray band. 

A second important result arises in {\it XMM} data from 1H\,0707-495. The source 
shows lags that oscillate between positive and negative values as a function of temporal frequency of 
variation. However, in 1H\,0707-495  it has also been shown \citep{miller10b} that 
a simple reverberation model can  explain the 
full set of (positive and negative) lags, 
requiring no additional physical processes to  explain the observed lag sign changes. 

Many AGN show a lag of hard X-ray photons relative to soft, with the
lag increasing to decreasing temporal frequency
\citep[e.g.][]{mchardy04a,papadakis01a,markowitz05a,arevalo06a,dasgupta06a,markowitz07a,sriram09a}. However,
NGC\,3516 shows no such positive lag (Figure~\ref{fig:lags}). In the context of our
reverberation model for AGN lags \citep{miller10a,miller10b}, the lack
of a lag in NGC\,3516 is interpreted as an absence of a  reverberation
signal.

These unusual lag results may be associated with the special sight-line at which we observe NGC\,3516. 
The results may be explained  
if the variations observed during 2006 are attributable 
predominantly to covering changes of the absorber along our line of sight (i.e. the
continuum being intrinsically quasi-constant during those observations), 
as we would not then expect to see 
systematic reverberation, which require all lines of sight to vary coherently.  
Even if the reprocessor subtended a
large solid angle to the continuum, there would be no reverberation for the case where there is 
no intrinsic variation in the continuum.  In this case,
the PIN-band fluctuations might be attributed to variable
obscuration by material having $N_{\rm H} > 10^{25}{\rm cm^{-2}}$.

The tentative negative lag observed at
low frequencies in NGC\,3516 might be explained by the presence of a
reprocessed signal in the soft-band, rather than in the hard band.  Such a soft signal would be
expected if we observed the emission spectrum from Compton-thin, ionized layers of the absorber. 
This suggestion is supported by the presence of significant 
soft-band line emission at the lowest X-ray flux states.

\section{Conclusions}

Comparison of 2009 {\it Suzaku} data with historic X-ray observations
of NGC\,3516 shows that relatively hard X-ray spectral forms are exhibited 
 at some epochs of high X-ray flux. Variable X-ray 
absorption is necessary to explain the range of states observed: simple changes in 
gas ionization state are insufficient to explain the range of properties exhibited by NGC\,3516. 

While much of the variability below 10\,keV may be explained by changes in the 
modeled X-ray absorber complex, there is a colorless component of variability 
evident during  high
flux states. In past work, the colorless flux variations have been suggested  to originate as
either intrinsic continuum fluctuations, or to be a consequence of the
passage of Compton-thick clumps of gas across the line-of-sight. 

NGC\,3516 provides a rare example to date of 
a source showing no significant lag of hard X-ray photons relative to soft photons 
over the timescales studied. This result may
be interpreted as an absence of any hard reverberation signal.
If this interpretation is correct, despite a
large amount of reprocessing material apparently being present around the
nucleus, then it seems likely that the continuum source is not intrinsically
strongly variable on the timescales studied. We suggest that even the high-state
flux variations may be  attributed to Compton-thick 
clumps crossing the line-of-sight.  The presence of such high-opacity 
clumps is a natural extension of the complex absorption model that can describe the 
X-ray spectrum. 

The unusual properties of NGC\,3516 may be a consequence of us viewing
the source at a rare orientation, where our line-of-sight intercepts
the edge of a significant absorber structure that collimates the
nuclear radiation leading to the observed biconical structure 
for the optical narrow-line-region gas. 

\acknowledgments

TJT acknowledges NASA grant NNX08AL50G. 
This research has made use of data obtained from the 
High Energy Astrophysics Science 
Archive Research Center (HEASARC), 
provided by NASA's Goddard Space Flight Center.



\bibliographystyle{apj}      
\bibliography{xray}   

\begin{deluxetable}{llll}
\tabletypesize{\scriptsize}
\rotate
\tablecaption{Observed X-ray Flux}
\tablewidth{0pt}
\tablehead{
\colhead{Observation} & 
\colhead{Flux (0.5-2)\,keV} &
\colhead{Flux (2-10)\,keV}  &
\colhead{Flux (10-50)\,keV} \\
} 
\startdata
2001-04 {\it XMM} & 0.49 & 2.27 & --  \\
2001-11 {\it XMM} & 0.29 & 1.63 &  -- \\
2005-10 {\it Suzaku} & 0.13   & 2.30 & 8.11 \\
2006-10 {\it XMM}/Mean & 1.75 & 4.36 & -- \\
2006-10 {\it XMM}/0401 & 2.21 & 5.08 & --  \\
2006-10 {\it XMM}/0501 &  2.02 & 4.49 & --   \\
2006-10 {\it XMM}/0601 &  1.04 & 3.62 & --   \\
2006-10 {\it XMM}/1001 &  1.95 & 4.44 & --   \\
2009-10 {\it Suzaku} & 0.30 & 1.30  &  3.90 \\ 
\enddata
\tablecomments{The observed flux in erg\,cm$^{-2}{\rm s^{-1}}$}. 
\end{deluxetable}

\begin{deluxetable}{lcccc}
\tabletypesize{\scriptsize}
\rotate
\tablecaption{Partial Covering Model} 
\tablewidth{0pt}
\tablehead{
\colhead{zone} &
\colhead{Log\,$\xi$} & 
\colhead{$N_{\rm H}$ \tablenotemark{1}} & 
\colhead{$C$ (2005)\tablenotemark{2}} &
\colhead{$C$ (2009)\tablenotemark{2}} \\ 
} 
\startdata

1a &  0.17$\pm0.28$  &  0.48$^{+0.18}_{-0.16}$ & $99\pm1$\%   &  $97\pm3$\%    \\
1b & 0.51$^{+0.25}_{-0.10}$ & 1.57$^{+0.64}_{-0.17}$ & -- & -- \\ 
2 & 2.00$^{+0.01}_{-0.02}$ & 19.0$^{+0.48}_{-0.61}$  & $99\pm1$\%& $86\pm1$\%   \\
3 & 2.05$\pm0.05$ & 200$\pm6$ & $70\pm2$ \% & $72^{+5}_{-2}$\% \\
\enddata
\tablecomments{A column of neutral gas covered all components, representing 
the Galactic absorption, see text for details.  An additional 
column of ionized gas covered all components during 2005 only, with 
$N_{\rm H} = 3 \times 10^{23}{\rm cm^{-2}}$, log\,$\xi =4.3$, see text for additional details.
 Errors are calculated at 90\% confidence. The remaining 
flux not accounted for in the Table, travels along a line-of-sight that is not 
covered by the intrinsic absorbers.} 
\tablenotetext{1}{Column density in units of 10$^{22} {\rm atom\, cm^{-2}}$}
\tablenotetext{2}{Percentage Covering}
\end{deluxetable}

\begin{deluxetable}{ll}
\tabletypesize{\scriptsize}
\rotate
\tablecaption{Fe\,K Emission Line}
\tablewidth{0pt}
\tablehead{
\colhead{Observation} & 
\colhead{Normalization\tablenotemark{1}} \\
} 
\startdata
2001-04 {\it XMM} & $3.33^{+0.40}_{-0.34}$   \\
2001-11 {\it XMM} & $4.02^{+0.25}_{-0.23}$   \\
2005-10 {\it Suzaku} & $5.75\pm 0.24$   \\
2006-10 {\it XMM}/Mean & $3.94\pm 0.17$   \\
2006-10 {\it XMM}/0401 & $3.51^{+0.67}_{-0.73}$   \\
2006-10 {\it XMM}/0501 & $4.00^{+0.61}_{-1.05}$   \\
2006-10 {\it XMM}/0601 & $4.25^{+0.47}_{-0.83}$   \\
2006-10 {\it XMM}/1001 & $3.96^{+0.82}_{-0.40}$   \\
2009-10 {\it Suzaku} & $4.16^{+0.33}_{-0.32}$   \\ 
\enddata
\tablecomments{For these fits the line energy was fixed at 6.4\,keV and the line width at $\sigma=40$\,eV, as 
determined from HEG data \citep{turner08a} 
 Errors were calculated 
at 90\% confidence}
\tablenotetext{1}{Normalization of a Gaussian model fit to the Fe\,K$\alpha$ line,
 in units 10$^{-5} {\rm photons\, cm^{-2}} {\rm s}^{-1}$. }
\end{deluxetable}

\end{document}